\documentclass{elsart}
\usepackage{latexsym,amssymb,amsmath,amsfonts,epsfig}

\makeatletter \@addtoreset{equation}{section} \makeatother
\renewcommand{\theequation}{\thesection.\arabic{equation}}
\newcommand {\rhs}       {right-hand side}
\newcommand {\bcs}       {boundary conditions}
\newcommand {\CS}        {Chern--Simons}
\newcommand {\chlgth}    {chiral lattice gauge theory}

\newcommand {\gsim}{\mathrel{\hbox{\rlap{\lower.55ex \hbox {$\sim$}}
            \kern-.3em \raise.4ex \hbox{$>$}}}}
\newcommand {\lsim}{\mathrel{\hbox{\rlap{\lower.55ex \hbox {$\sim$}}
            \kern-.3em \raise.4ex \hbox{$<$}}}}

\newcommand {\beq}[0]{\begin{equation}}
\newcommand {\eeq}[0]{\end{equation}}
\newcommand {\nn}    {\nonumber}
\renewcommand {\bar}     {\overline}

\def\R{\mathbb{R}}
\def\Z{\mathbb{Z}}
\def\d{{\rm d}}
\def\e{{\rm e}}

\def\id{\makebox[0.6ex][l]{$1$}{\rm l}}

\def\1{\id}                             %%without bbm.sty

\def\ncirc{\overset{\circ}{n}}
\def\nstar{\overset{\star}{n}}

\begin{document}
\noindent
Nuclear Physics B 639 (2002) 241  \hspace*{\fill} hep-th/0205038  \newline
                                  \hspace*{\fill} KA--TP--08--2002\newline
\vspace*{1cm}
\runauthor{Klinkhamer and Schimmel}
\begin{frontmatter}

\title{CPT anomaly: a rigorous result in four dimensions}

\author{F.R.~Klinkhamer},
\ead{frans.klinkhamer@physik.uni-karlsruhe.de}
\author{J.~Schimmel}
\ead{schimmel@particle.uni-karlsruhe.de}

\address{Institut f\"ur Theoretische Physik, Universit\"at
Karlsruhe, D--76128 Karlsruhe, Germany}

\begin{abstract}
The existence of a CPT anomaly is established for a particular
four-dimensional Abelian lattice gauge theory with Ginsparg-Wilson fermions.
\end{abstract}

\end{frontmatter}

\section{Introduction}
Recently, it has been claimed that certain four-dimensional chiral gauge
theories defined over a topologically nontrivial spacetime manifold (e.g.,
$M = \R \times S^1 \times S^1 \times S^1$) display an anomalous breaking of
Lorentz and time-reversal (T) invariance. Since the combined CPT
invariance would also be broken (C stands for charge conjugation and P
for parity reflection), the effect has been called a CPT
anomaly~\cite{K00}.  See Ref.~\cite{K01} for a review and further references.

A related effect has been found in a class of exactly solvable $U(1)$
gauge theories in two spacetime dimensions~\cite{KN01,KM01}.
In two dimensions, the CPT anomaly is well established,
because the ``chiral'' determinant is known explicitly.
But there is, of course, no exact result
for the chiral determinant in four dimensions. Still, we can try to make
sure of the existence of the four-dimensional CPT anomaly
by carefully studying one particular case.

We, therefore, consider in this paper an Abelian chiral gauge theory
over the spacetime manifold $M = \R^3 \times S^1$, with trivial
vierbeins (tetrads), $e^a_\mu (x)= \delta^a_\mu$,
and a corresponding flat metric of Euclidean
signature, $g_{\mu\nu} (x) = \delta_{\mu\nu}$.
[The Kronecker symbols $\delta^a_\mu$ and $\delta_{\mu\nu}$,
with indices running over 1, 2, 3, 4, take the value 1 if
their respective indices are equal and the value 0 otherwise.]
The spinors are taken to
be periodic over the compact dimension, i.e., there is a periodic spin
structure. Specifically, the gauge group $G$, the representation $R_L$
of the left-handed fermions, the spacetime manifold $M$ and the vierbeins
$e^a_\mu(x)$ are given by
\begin{eqnarray}
\left[ \, G;\; R_L;\; M;\; e^a_\mu \,\right] =
\left[\, U(1);\; 8 \times (+1) +
       1 \times (-2);\; \R^3 \times S^1_\mathrm{\,PSS};\; \delta^a_\mu \,\right].
\label{theory}
\end{eqnarray}
Here, $R_L$ determines the electric charges of the left-handed fermions
(in terms of a fundamental unit)
and PSS stands for the periodic spin structure.
The Abelian chiral gauge theory (\ref{theory}) has $N_F=9$ Weyl
fermions, with the perturbative gauge anomaly cancelling out
\cite{A69,BJ69,BIM72,GJ72}.

The ultraviolet divergences of the Abelian chiral gauge theory
(\ref{theory}) are regulated by the introduction of a hypercubic
spacetime lattice.  In addition, so-called Ginsparg-Wilson fermions~\cite{GW82}
are used, with the explicit lattice Dirac operator found by Neuberger~\cite{N98}.
The corresponding Euclidean chiral $U(1)$ gauge theory on the
lattice has been formulated by L{\"u}scher~\cite{L99} and we refer to his paper
for further details.

It is our goal to establish the CPT anomaly of this specific regularized
theory. For our purpose, it is important that the nonperturbative
regularization method used respects both gauge invariance and a lattice
version of chiral symmetry. This new calculation should also be an
improvement over the previous one in Ref.~\cite{K00}, which was perturbative
and which used an \emph{ad hoc} ultraviolet regularization. Moreover, finding
the same result with two different regularization methods would suggest
that the four-dimensional CPT anomaly is not simply an artifact of a
particular regularization.

The outline of this paper is as follows. In Section~\ref{general}, we
describe the setup of the calculation for the lattice version of the
Abelian chiral gauge theory (\ref{theory}). Specifically, we restrict
ourselves to gauge fields which are independent of the compact coordinate
and have trivial holonomies over the compact dimension. In
Section~\ref{ChiralLattice}, we recall the basic points of chiral
lattice gauge theory, but refer to Ref.~\cite{L99} for further details.
(Some matters of notation are relegated to Appendix~\ref{Notation}.)

In Section~\ref{EffeCPT}, we discuss the resulting effective gauge field
action and its behavior under a CPT transformation
of the gauge field configuration considered.
In Section~\ref{oddN}, the effective action
is found to change under CPT, for arbitrary lattice spacing $a$ and
arbitrary odd integer $N = L/a$ (where $L$ is the size of the periodic
compact dimension). This establishes the CPT anomaly for an odd number
$N$ of links in the periodic direction, which is the main result of our
paper. (Appendix~\ref{Fermionicmeasure} deals with a technical point
regarding the definition of the fermion integration measure.)

In Section~\ref{evenN}, we briefly discuss the
case of even $N$, which reproduces the previous result for odd $N$ in
the classical continuum limit. In Section~\ref{Ntwo}, we calculate the $N=2$
effective action for smooth gauge field backgrounds in the continuum
limit and the CPT anomaly is manifest.

In Section~\ref{Discussion},
finally, we summarize our results and raise a question about reflection
positivity.

\section{General setting}\label{general}
In this section, we present the lattice version of the Abelian gauge
theory (\ref{theory}) and the setup for the calculation to follow.  In
the next section, we discuss the specific issues relevant to chiral
symmetry on the lattice.

In order to regulate possible infrared divergences, we temporarily
consider a finite spacetime volume
\begin{eqnarray}
V_4 = V^\prime \times L = (L')^3 \times L, \label{volume}
\end{eqnarray}
with lengths $L$ and $L' \gg L$. Ultraviolet divergences are
regulated by the introduction of a regular hypercubic lattice:
\begin{eqnarray}
L' = N'\, a, \qquad L = N\, a, \label{lengths}
\end{eqnarray}
with lattice spacing $a$
and positive integers $N$ and $N' \gg N$. The points of the lattice have
Euclidean spacetime coordinates
\begin{eqnarray}
(x_1,x_2,x_3,x_4) \equiv (\vec x, x_4) = (\vec n \, a, n_4\, a),
\end{eqnarray}
with integers
\begin{eqnarray}
n_1,n_2,n_3 \in [0,N'\,], \qquad n_4 \in [0,N\,].
 \label{n1234}
\end{eqnarray}

Furthermore, we let the Euclidean spacetime
indices $\kappa$, $\lambda$, $\mu$, $\ldots$, run over the coordinate labels
1, 2, 3, 4, and the spacetime indices $k$, $l$, $m$, $\ldots$, over 1, 2, 3.
Note also that one of the coordinates of $\vec x$, say $x_1$, should
correspond to the time coordinate after the Wick rotation, whereas the other
coordinates $x_2$, $x_3$, and $x_4$, should become spacelike
(see Ref.~\cite{K00} and Section \ref{Discussion} below).

As usual in $U(1)$ lattice gauge theory, the spinor fields $\psi_f$,
with flavor index $f=1,\ldots,N_F$,
reside at the sites of the hypercubic lattice and the variable
$U_\mu(x) \in U(1)$ is associated with the directed link between the site
$x$ and its nearest neighbor in the $\mu$ direction.
The lattice fields over the volume
(\ref{volume}) have periodic boundary conditions in $x_4$:
\begin{eqnarray}
\psi_f (\vec x, L) &=&+ \psi_f(\vec x, 0), \quad
\overline \psi_f (\vec x, L) = +\overline \psi_f(\vec x, 0),\nn \\[0.2cm]
U_\mu (\vec x, L) &=& +U_\mu (\vec x, 0). \label{pbcs}
\end{eqnarray}
For the other coordinates $\vec x$, the link variables are again periodic
but the fermion fields are taken to be antiperiodic:
\begin{eqnarray}
\psi_f (L', x_2, x_3,x_4) &=& -\psi_f(0, x_2,x_3,x_4),
\nonumber\\[0.2cm]
\overline \psi_f (L',x_2,x_3,x_4) &=& -\overline \psi_f (0,x_2,x_3,x_4),
\nonumber\\[0.2cm]
U_\mu (L', x_2, x_3,x_4) &=& + U_\mu (0, x_2, x_3,x_4),
\label{apbc}
\end{eqnarray}
and similarly for $x_2$ and $x_3$. These antiperiodic boundary
conditions provide an infrared regulator for the massless fermions, as will
become clear later.

As explained in Ref.~\cite{K00}, it suffices to establish the
CPT anomaly for one particular configuration of the classical gauge
fields. Concretely, this will be done by showing the noninvariance of the
effective action, to be defined in Section~\ref{Effe}.
To simplify the calculation, the link variables in the $4$ direction
are taken to be 1 and the link variables in the other three
directions to be $x_4$-independent:
\begin{eqnarray}
U_4 (\vec x, x_4) = 1, \qquad U_m (\vec x, x_4) = U_m (\vec x), \quad m
=1,2,3. \label{config}
\end{eqnarray}
In addition, we set the link variables $U_m(\vec x)$
to 1
if they belong to a plaquette on the boundary of the volume $V'$.
This implies that all flux quantum numbers $m_{\mu\nu}$ vanish;
cf. Ref.~\cite{L99}.

For these particular link variables, the holonomies in the
4 direction are trivial,
\begin{eqnarray}
h(\vec x)  \equiv \prod_{n_4=0}^{N-1}\,U_4 (\vec x, n_4\, a) = 1.
\label{holonomy}
\end{eqnarray}
The holonomies are, of course, invariant under periodic gauge transformations.
Furthermore, the real continuum gauge field $A_\mu(x)$ is related to the
link variable in the usual way:
\begin{eqnarray} \label{expieaA}
U_\mu (x) = \exp \left( i e \int\limits_{x}^{x+a \hat{\mu}} \d y^{\nu} A_\nu(y)
\right) \approx \exp \Big( i e a A_\mu(x+ a \hat{\mu}/2) \Big),
\end{eqnarray}
with the dimensionless coupling constant $e$ and
the unit vector $\hat{\mu}$ in the $\mu$ direction.
The approximate equality on the right of
Eq.~(\ref{expieaA}) holds in the classical continuum limit
$a \to 0$; see, e.g., Ref.~\cite{MM94} for further details.

\section{Chiral lattice gauge theory}\label{ChiralLattice}
It is well known that lattice gauge theory, in its simplest form, displays
an unwanted doubling of fermions species if the spinor fields are
defined at the sites of the hypercubic lattice; cf. Ref. \cite{MM94}.
This problem can be circumvented
by the introduction of a term of second order in the difference operator.
The resulting Wilson-Dirac operator,
\begin{eqnarray}
D_{\rm W} = \half \sum_{\mu=1}^4 \Big( \gamma_\mu \big(\nabla_\mu +
  \nabla_\mu^* \big) - a\nabla_\mu^*\nabla_\mu \Big),
  \label{WilsonDirac}
\end{eqnarray}
breaks, however, chiral symmetry (see Appendix~\ref{Notation} for notation).

To effectively restore the chiral symmetry, one can use a new
gauge-covariant operator $D$ which satisfies the Ginsparg-Wilson
relation~\cite{GW82}
\begin{eqnarray}\label{GWrelation}
\gamma_5 D + D \gamma_5 = a D \gamma_5 D,
\end{eqnarray}
with lattice spacing $a$ and chirality matrix $\gamma_5$.
This operator may be of the form
\begin{eqnarray}
D[U] = a^{-1}\, \big(\1 - V[U] \big), \label{GWop}
\end{eqnarray}
with a unitary operator $V$ that depends on the gauge field
configuration $U_\mu(x)$.
In order to satisfy the Ginsparg-Wilson relation,
$V$ has to be $\gamma_5$-Hermitian:
\begin{eqnarray}\label{Vdagger}
\gamma_5 V \gamma_5 = V^\dagger.
\end{eqnarray}
An explicit example of the operator $V$ is known \cite{N98}:
\begin{eqnarray}
V = X\left(X^\dagger X\right)^{-1/2} = \int\limits_{-\infty}^\infty
\frac{\d t}{\pi}\; X \left(t^2 + X^\dagger X\right)^{-1}, \label{unitop}
\end{eqnarray}
with
\begin{eqnarray}\label{X}
X \equiv \1 - a D_{\rm W},
\end{eqnarray}
for the Wilson-Dirac operator $D_{\rm W}$
defined by Eq.~(\ref{WilsonDirac}). But, for the moment, the operator
$V$ is kept general.

The lattice fermion action \cite{L99}
\begin{eqnarray}
S_F = a^4 \sum_x \,\sum_{f=1}^{N_F}\,
\overline \psi_f (x)\, D[U]\, \psi_f(x) \label{action}
\end{eqnarray}
is then invariant under the following infinitesimal transformation:
\begin{eqnarray}
\psi_f(x) \to \psi_f (x) + \delta \psi_f(x) , \qquad
\overline\psi_f(x) \to \overline \psi_f (x) + \delta \overline \psi_f(x),
\end{eqnarray}
with
\begin{eqnarray}
\delta \psi_f (x) &=& i\varepsilon\, \gamma_5 V \psi_f(x) \equiv
i\varepsilon\, \hat \gamma_5 \psi_f(x) ,\quad
\delta \overline \psi_f(x) = i\varepsilon\, \overline \psi_f(x) \gamma_5 ,
\end{eqnarray}
for an infinitesimal parameter $\varepsilon$. Note that for $V=\1$ one
recovers the usual chiral transformation.

A chiral lattice gauge theory
can now be constructed by imposing the constraints~\cite{L99}
\begin{eqnarray}
\psi_f(x) = \hat P_- \, \psi_f(x), \qquad
\overline \psi_f(x) = \overline \psi_f(x)\, P_+,
\label{constrain}
\end{eqnarray}
with the projection operators (recall $\hat\gamma_5 \equiv \gamma_5 V$)
\begin{eqnarray}
\hat P_\pm \equiv \half (\1 \pm \hat \gamma_5) , \qquad P_\pm \equiv
\half (\1 \pm \gamma_5). \label{project}
\end{eqnarray}

For the calculations of this paper
we use the explicit solution  (\ref{GWop})--(\ref{X}) of the Ginsparg-Wilson
relation (\ref{GWrelation}).
The fermion action is then given by Eq.~(\ref{action}),
the chirality constraints by Eq.~(\ref{constrain}), and the \bcs~by
Eqs.~(\ref{pbcs})--(\ref{apbc}).
Unless stated otherwise, we work in the following with a single fermion of
unit charge and drop the flavor index $f$.

\section{Effective action and CPT transformation}\label{EffeCPT}
\subsection{Effective action}\label{Effe}
The effective gauge field action is obtained by integrating out the
fermionic degrees of freedom.
For the lattice gauge theory of the previous two sections, the effective action
$\Gamma[U]$ is given by the following integral:
\begin{eqnarray}
\exp\left(- \Gamma[U]\right) &=&
K \int \prod_x \d \psi(x) \prod_x \d \overline \psi(x) \; \exp\left(
  -S_F[\overline \psi, \psi, U] \right), \label{effact}
\end{eqnarray}
with the action $S_F$ as defined by Eqs.~(\ref{GWop})--(\ref{action}) for
$N_F = 1$ and fermion fields that satisfy the chirality constraints
(\ref{constrain}) and \bcs~(\ref{pbcs})--(\ref{apbc}).
The normalization constant $K$ assures that $\Gamma[1]=0$
for the constant gauge field configuration $U_\mu (x)=1$.

Because of the periodic boundary conditions (\ref{pbcs}), the
fermions can be decomposed into a finite series of Fourier modes,
\begin{eqnarray}
\psi(x) &=& \sum_n \xi_n (\vec x) \, \e^{2\pi i n x_4 /L}, \quad
\overline \psi(x) = \sum_n \overline \xi_n (\vec x) \,
\e^{-2\pi i n x_4 /L}, \label{fermimod}
\end{eqnarray}
where the integer $n$ takes the values
\begin{eqnarray}
  -(N-1)/2     \leqslant n \leqslant (N-1)/2,     \quad \mbox{for odd $N$},
  \label{nodd}
\end{eqnarray}
and
\begin{eqnarray}
  -(N/2) +1    \leqslant n \leqslant (N/2),    \quad \mbox{for even $N$},
 \label{neven}
\end{eqnarray}
with $N \equiv L/a$ the number of links in the 4 direction.

For the gauge field configurations (\ref{config}), the action of the
operator (\ref{unitop}) on the fermion field (\ref{fermimod}) is
\begin{eqnarray}
V \psi (x) &=& V \sum_n \xi_n (\vec  x)\, \e^{2\pi i n x_4 / L} \nonumber\\
&=& \sum_n \e^{2\pi i n x_4 / L} \int\limits_{-\infty}^\infty \frac{\d
  t}{\pi}\; X^{(n)} \left(t^2 + X^{(n)}{}^\dagger X^{(n)}
  \right)^{-1} \xi_n (\vec x) \nonumber\\
&\equiv& \sum_n \e^{2\pi i n x_4 / L}\, V^{(n)}\, \xi_n(\vec x).\label{Vn}
\end{eqnarray}
Here, the operators $X^{(n)}$ are defined by
\begin{eqnarray}
X^{(n)} &\equiv&
\1 - \left(a D_\mathrm{\,W}^\mathrm{\,3D} + i \gamma_4 \ncirc - (\nstar - \1)
\right) = \nstar - aD_\mathrm{\,W}^\mathrm{\,3D} - i \gamma_4 \ncirc, \label{Xn}
\end{eqnarray}
with the ``three-dimensional'' Wilson-Dirac operator
\begin{eqnarray}\label{DW3dim}
D_\mathrm{\,W}^\mathrm{\,3D} =
\half \sum_{\mu=1}^3 \Big( \gamma_\mu \big(\nabla_\mu +
  \nabla_\mu^*\big) - a\nabla_\mu^*\nabla_\mu \Big)
\end{eqnarray}
and the further notation
\begin{eqnarray}
\nstar \equiv \cos \left(2\pi n/N \right), \qquad \ncirc \equiv \sin
\left( 2\pi n/N \right).
\end{eqnarray}
Note that the operator (\ref{DW3dim}) is still defined with the $4\times 4$
matrices $\gamma_\mu$ given by Eq.~(\ref{gammamu}) of Appendix~\ref{Notation}.

In terms of the Fourier modes $\xi_n$  and $\overline \xi_n$ of
Eq.~(\ref{fermimod}), the action $S_F$ becomes
\begin{eqnarray}
S_F = \sum_n \left( a^3 \sum_{\vec x} N\; \overline \xi_n(\vec x)\,
  \big(\1 - V^{(n)}[U]\big)\, \xi_n(\vec x) \right).
\end{eqnarray}
Redefining the fermion fields
\begin{eqnarray}
\chi_n (\vec x) \equiv N^{1/2} \; \xi_n(\vec x) , \qquad
\overline \chi_n (\vec x) \equiv N^{1/2} \;\overline \xi_n (\vec x) , \label{fermi}
\end{eqnarray}
the action reads
\begin{eqnarray}
S_F &=& \sum_n \left( a^3 \sum_{\vec x}
  \overline \chi_n (\vec x)\, a D^{(n)}[U]\, \chi_n (\vec x) \right) \equiv
  \sum_n S_F^{(n)} [\overline \chi_n, \chi_n, U] , \label{fourier_action}
\end{eqnarray}
with
\begin{eqnarray}\label{Dn}
D^{(n)}[U] \equiv a^{-1}\, \left(\1 - V^{(n)}[U]\right).
\end{eqnarray}

As mentioned above, the fermions (\ref{fermimod})
must satisfy the constraints (\ref{constrain}). This implies that the modes
$\chi_n$ and $\overline \chi_n$ have corresponding constraints,
\begin{eqnarray}\label{chinprojection}
\chi_n (\vec x) = \hat P^{(n)}_- \, \chi_n (\vec x) , \qquad \overline
\chi_n (\vec x) = \overline \chi_n (\vec x)\, P_+, \label{constrain_n}
\end{eqnarray}
with the projection operator $P_+$ defined by Eq.~(\ref{project}) and the
$n$-dependent projection operators
\begin{eqnarray}\label{Pminhatn}
\hat P_-^{(n)}[U] \equiv \half \left(\1 - \gamma_5 V^{(n)}[U]\right) \equiv
\half \left(\1 - \hat \gamma_5^{(n)}[U] \right).
\end{eqnarray}
The operators $\hat \gamma_5^{(n)}$ are Hermitian unitary operators and
can be written as follows:
\begin{eqnarray}\label{gamma5hatn}
\hat \gamma_5^{(n)} [U] = \left(
\begin{array}{cc}
h_1^{(n)}[U]               &\;\; i h_2^{(n)} [U] \\
-i h_2^{(n)}{}^\dagger [U] &\;\; h_3^{(n)} [U]
\end{array}
\right) ,
\end{eqnarray}
with Hermitian operators $h_1^{(n)}$ and $h_3^{(n)}$. The
unitarity of $\hat \gamma_5^{(n)}$ gives then
\begin{eqnarray}
\left(h_1^{(n)} \right)^2+ h_2^{(n)} \, h_2^{(n)}{}^\dagger  &=&
\1, \quad
\left(h_3^{(n)} \right)^2 + h_2^{(n)}{}^\dagger \, h_2^{(n)}  =
\1, \nn\\[0.2cm]
 h_1^{(n)} \, h_2^{(n)}  + h_2^{(n)} \, h_3^{(n)}  &=& 0.
\end{eqnarray}

The crucial observation, now, is that the fermionic integral (\ref{effact})
\emph{factorizes} for the gauge field configurations (\ref{config}):
\begin{eqnarray}\label{factorization}
\!\!\! \exp\left(-\Gamma[U]\right) &=&
K^{\prime}\,\prod_n \int
\prod_{\vec x} \d \chi_n(\vec x) \prod_{\vec x} \d \overline \chi_n(\vec x)\;
\exp\left(-S_F^{(n)}[\overline \chi_n, \chi_n, U]\right),
\end{eqnarray}
with the Fourier index $n$ running over the values (\ref{nodd}) or
(\ref{neven}), the actions $S_F^{(n)}$ defined by Eq.~(\ref{fourier_action})
and the implicit constraints (\ref{constrain_n}).
[It is instructive to compare the finite product (\ref{factorization})
with the infinite products (3.7) and (3.9) of Ref.~\cite{K00},
which were regulated by an infinite tower of Pauli--Villars masses.]

In order to make the fermionic integrations in Eq.~(\ref{factorization})
more explicit, the three-dimensional fermion fields (\ref{fermi})
are expanded as follows:
\begin{eqnarray}
\chi_n (\vec x) = \sum_j v_j^{(n)} (\vec x)\, c_j^{(n)}, \qquad \overline
\chi_n(\vec x) = \sum_k \bar c_k^{(n)}\, \bar v_k^{(n)}(\vec x).
\label{bases}
\end{eqnarray}
Here, the coefficients $c_j^{(n)}$, for $j= 1,2,3,\ldots$,  and
$\bar c_k^{(n)}$, for $k=1,2,3,\ldots$,
are anticommuting (Grassmann) numbers (cf. Ref.~\cite{MM94}).
The $v_j^{(n)}(\vec x)$ and $\bar v_k^{(n)}(\vec x)$ build
complete orthonormal bases of lattice
spinors, which satisfy the constraints (\ref{constrain_n}) and have inner products
\begin{eqnarray}
\left(v_i^{(n)},v_j^{(n)}\right) &=& a^3 \sum_{\vec x} v_i^{(n)}{}^\dagger(\vec x)\,
 v_j^{(n)}(\vec x) = \delta_{ij},\nn\\
\left(\bar v_k^{(n)}{}^\dagger,\bar v_l^{(n)}{}^\dagger\right) &=& a^3 \sum_{\vec
 x} \bar v_k^{(n)}(\vec x)\, \bar v_l^{(n)}{}^\dagger (\vec x) =
 \delta_{kl}.
\end{eqnarray}
Note that the spinor $v_j^{(n)}$ depends on the link variables $U$, because
the projection operator $\hat P_-^{(n)}$ in Eq.~(\ref{chinprojection})
depends on the gauge field configuration. In the following, this dependence will
be emphasized by writing $v_j^{(n)} (\vec x; U)$.

With these expansions, the effective action (\ref{factorization}) for the gauge
field configurations (\ref{config}) is given by the following fermionic integral:
\begin{eqnarray}
\exp\left(-\Gamma[U]\right) &=&
K^{\prime\prime}\, \prod_n \int \prod_j \d c_j^{(n)}
\prod_k \d \bar c_k^{(n)}\: \exp \left( - \sum_{k,j} \bar c_k^{(n)}\,
  M_{kj}^{(n)}[U]\, c_j^{(n)} \right),\nn\\
&& \label{integral}
\end{eqnarray}
with the matrices
\begin{eqnarray}
M_{kj}^{(n)} [U] \equiv a^3 \sum_{\vec x} \bar v_k^{(n)} (\vec x)\, a D^{(n)}[U] \,
v_j^{(n)} (\vec x; U), \label{matrix}
\end{eqnarray}
and $D^{(n)}$ defined by Eq.~(\ref{Dn}). The constant $K^{\prime\prime}$
normalizes the integral (\ref{integral}), so that $\Gamma[1]=0$.

\subsection{CPT transformation}\label{measure_change}
The goal of this paper is to investigate the behavior of the effective action
(\ref{integral}) under a CPT transformation of the gauge field.
First, recall that, according to Eq.~(\ref{config}),
the gauge fields considered are $x_4$ independent. Let $\mathcal{R}$ now be
the coordinate reflection operator in the remaining three dimensions,
\begin{eqnarray}\label{reflection}
\mathcal{R} : \vec x \to -\vec x.
\end{eqnarray}
The three-dimensional Wilson-Dirac operator (\ref{DW3dim})
has then the following behavior under a CPT transformation:
\begin{eqnarray}
\mathcal{R}\, D^\mathrm{\,3D}_\mathrm{\,W}\,[U]\, \mathcal{R} =
D^\mathrm{\,3D\,\dagger}_\mathrm{\,W}\,[U^\theta],
\end{eqnarray}
with the CPT-transformed link variable
\begin{eqnarray} \label{Utheta}
U^\theta_m (\vec x) \equiv U^\dagger_m (-\vec x -a\,\hat{m}),\qquad m=1,2,3.
\end{eqnarray}
(In our case, we have also $U^\theta_4 = U_4 = 1$.)
This implies that the Fourier modes $ D^{(n)}$ of the Ginsparg-Wilson
operator $D$ transform as follows:
\begin{eqnarray}
\mathcal{R}\, D^{(n)}\,[U]\, \mathcal{R} &=& D^{(-n)}{}^\dagger\, [U^\theta].
\end{eqnarray}
Equivalently, one has
\begin{eqnarray}
\mathcal{R}\, V^{(n)}\,[U]\, \mathcal{R} &=& V^{(-n)}{}^\dagger\, [U^\theta].
\label{cptV}
\end{eqnarray}

The matrices $M_{kj}^{(n)} [U]$, which enter the fermionic integral
(\ref{integral}),  change under the substitution $U \to U^\theta$ to
\begin{eqnarray}\label{MUtheta}
M_{kj}^{(n)} [U^\theta] &=& a^4 \sum_{\vec x} \bar v_k^{(n)} (\vec x)\,
D^{(n)}[U^\theta] \, v_j^{(n)} (\vec x; U^\theta) \nonumber\\
&=& a^4 \sum_{\vec x} \bar v_k^{(n)} (\vec x)\, \mathcal{R} \gamma_5 \,
D^{(-n)}[U]\, \gamma_5 \mathcal{R}\, v_j^{(n)} (\vec x; U^\theta)
\nonumber \\
&=& \sum_{l,i} \bar{\mathcal{Q}}^{(-n)}_{kl} \left( a^4 \sum_{\vec x} \bar
v_l^{(-n)} (\vec x)\, D^{(-n)}[U]\, v_i^{(-n)} (\vec x; U) \right)
\mathcal{Q}^{(-n)}_{ij} \nonumber \\
&=& \sum_{l,i}\bar{\mathcal{Q}}^{(-n)}_{kl}\, M^{(-n)}_{li} [U] \,
\mathcal{Q}^{(-n)}_{ij} ,
\end{eqnarray}
in terms of the unitary transformation matrices
\begin{eqnarray}\label{Qdefinition}
\mathcal{Q}^{(-n)}_{ij}[U] &\equiv& a^3 \sum_{\vec x}
v_i^{(-n)}{}^\dagger (\vec x; U)\, \gamma_5 \mathcal{R}\, v_j^{(n)}
(\vec x; U^\theta), \nn\\
\bar{\mathcal{Q}}^{(-n)}_{kl} &\equiv& a^3 \sum_{\vec x} \bar
v_k^{(n)}(\vec x)\, \gamma_5 \mathcal{R}\, \bar v_l^{(-n)}{}^\dagger
(\vec x).
\end{eqnarray}
For the second equality in Eq.~(\ref{MUtheta}), use has been made of the
$\gamma_5$-Hermiticity of $D$; cf. Eq.~(\ref{Vdagger}).
Since the gauge fields considered are topologically trivial
(see Eq.~(\ref{config}) and the sentences below it),
there is no fermion number violation \cite{L99}
and the matrices $\mathcal{Q}^{(-n)}$ and $\bar{\mathcal{Q}}^{(-n)}$
have equal dimensions.

The transformation matrices in the last expression of Eq.~(\ref{MUtheta})
can be absorbed by a redefinition of the fermionic variables in
the integral (\ref{integral}), but the measure picks up a Jacobian
(just as happens for the well-known chiral anomaly \cite{F80}).
This implies the following change of the effective action under a CPT
transformation of the gauge field configurations (\ref{config}):
\begin{eqnarray} \label{GammaUtheta}
\Gamma[U^\theta] = \Gamma[U] - \sum_n
\ln \det\left( \sum_{l\, }
\mathcal{Q}^{(n)}_{kl}[U]\,\bar{\mathcal{Q}}^{(n)}_{lm}
\right).\,
\end{eqnarray}
The determinants of the transformation matrices $\mathcal{Q}^{(n)}$
may, in principle, depend on the link variables $U_m(\vec x)$
and there is the possibility that the effective action is CPT noninvariant.
Any change of the effective action would, however, be purely
imaginary because of the unitarity of the transformation matrices
$\mathcal{Q}^{(n)}$ and $\bar{\mathcal{Q}}^{(n)}$.

\section{CPT anomaly for odd $N$}\label{oddN}
We have seen in Section~\ref{measure_change} that the integration measure
could change under a CPT transformation. But to calculate
explicitly the corresponding change of the
effective action, one needs to know the bases $v_j^{(n)}$,
$\bar v_k^{(n)}$ and how they transform under CPT;
cf. Eqs.~(\ref{Qdefinition}) and (\ref{GammaUtheta}).

\subsection{Preliminaries}\label{Preliminaries}
The basis spinors of the conjugated fermions are easy to find (the Dirac matrices
are given in Appendix A). The four-component basis spinors are simply
\begin{eqnarray} \label{phibar}
\bar v_k^{(n)}(\vec x) \equiv  \bar \varphi_k^{(n)}(\vec x)
= \big( \bar \psi_k (\vec x), 0 \big),
\end{eqnarray}
where the $\bar \psi_k$ build an arbitrary orthonormal basis of conjugated
two-spinors.  An example would be
\begin{eqnarray}\label{psibar}
\bar \psi_k(\vec x) \equiv \bar \psi_{a, k'_1, k'_2, k'_3}(\vec x) =
 (L')^{-3/2} \exp \left( -i\,\pi\,
                  (2\,\vec k^{\,\prime} + \vec b) \cdot \vec x /L' \right)
 \, \bar \psi_a,
\end{eqnarray}
with the constant two-spinors $\bar \psi_1 \equiv (1,0)$ and $\bar \psi_2
\equiv (0,1)$. Here, the vector $\vec b \equiv (1,1,1)$ assures
the antiperiodicity (\ref{apbc}) and $\vec k^{\,\prime}$ stands for a
triple of integers $(k'_1, k'_2, k'_3)$ with finite range, as in
Eqs.~(\ref{nodd})--(\ref{neven}) but with $N$ replaced by $N^{\,\prime}$.
The bases (\ref{phibar}) are chosen to be equal for all Fourier modes $n$.

The basis spinors $v_j^{(n)} (\vec x ;U)$ are more difficult to find. The
problem of determining these spinors can be shifted to the problem of finding
unitary operators $\tilde U^{(n)}[U]$ that diagonalize the operators
$\hat \gamma_5^{(n)}[U]$ (cf. Ref.~\cite{Na98}):
\begin{eqnarray}
\tilde U^{(n)}{}^\dagger[U] \; \hat \gamma_5^{(n)}[U]\; \tilde U^{(n)}[U] =
\gamma_5. \label{diag}
\end{eqnarray}
If these operators are found, the four-component basis spinors are given by
\begin{eqnarray}\label{vjn}
v_j^{(n)} (\vec x ;U) = \tilde U^{(n)}[U]\; \varphi_j(\vec x),
\end{eqnarray}
with an orthonormal basis $\varphi_j$ that satisfies the constraint
$P_-\, \varphi_j = \varphi_j$. This basis is  simply
\begin{eqnarray} \label{phi}
\varphi_j (\vec x) =
\left(
\begin{array}{c}
0\\ \psi_j(\vec x)
\end{array}
\right),
\end{eqnarray}
where the two-spinors $\psi_j(\vec x)$ can be defined analogously to
Eq.~(\ref{psibar}).

The operators $\tilde U^{(n)}$ are, however, not unique. Equation (\ref{diag})
is also satisfied with $\tilde U^{(n)}$ replaced by $\tilde U^{(n)} Y^{(n)}$,
for an arbitrary unitary matrix
\begin{eqnarray}
Y^{(n)} = \left(
\begin{array}{cc}
Y_1^{(n)} & 0 \\
0 & Y_2^{(n)}
\end{array}
\right), \qquad
Y_1^{(n)}{}^\dagger\: Y_1^{(n)} = \1 ,\quad
Y_2^{(n)}{}^\dagger\: Y_2^{(n)} = \1 .
\label{ambig}
\end{eqnarray}
There is thus a phase ambiguity in the integration measure \cite{L99},
which will be used later on.

We now consider the case of an odd number $N$ of links in the
4 direction (recall $N \equiv L/a$). The restriction to odd $N$ is
useful because the operators $\hat \gamma_5^{(n)}$, defined by
Eqs.~(\ref{Vn}) and (\ref{Pminhatn}), have the following property:
\begin{eqnarray}
\hat \gamma_5^{(n)} \tilde \gamma_4 = - \tilde \gamma_4 \hat
\gamma_5^{(-n)}, \label{prop1}
\end{eqnarray}
with the matrix\footnote{The crucial properties of $\tilde \gamma_4$
are its commutation with $\gamma_m$, for $m=1,2,3$,
and its anticommutation with $\gamma_4$. Recall that the $x_4$
coordinate is singled out by the \bcs~and that the vierbeins
are trivial; cf. Ref.~\cite{KM01}.}
\begin{eqnarray}\label{gamma4tilde}
\tilde \gamma_4 \equiv i \gamma_4 \gamma_5 = \left(
\begin{array}{cc}
0 &\;\; \1 \\ \1 &\;\; 0
\end{array}
\right).
\end{eqnarray}
Therefore, the set of operators $\hat \gamma_5^{(n)}$,
for $n$ in the range (\ref{nodd}) with $N \geq 3$,
splits into the two related subsets $n<0$ and $n>0$ and the separate subset
$n=0$. [For even $N$,  the Fourier index $n$ runs over the range (\ref{neven}) and
one has, in general, to consider four subsets: $-N/2<n<0$, $0<n<N/2$, $n=0$ and
$n = N/2$. The restriction to odd $N$ is purely
technical, at least in the continuum limit $a \to 0$, $N \to \infty$,
$N a = {\rm constant}$. See Section~\ref{evenN} for further details.]

Suppose that the operators $\tilde U^{(n)}$ are known for $n > 0$.
Then the $\tilde U^{(n)}$ for $n<0$ can be obtained as follows.
First, multiply Eq.~(\ref{diag}) with $\tilde \gamma_4$ from the right
and the left. Then, insert $\1 = \tilde \gamma_4 \tilde \gamma_4$
and use the relation (\ref{prop1}). This gives
\begin{eqnarray}
\tilde \gamma_4 \tilde U^{(n)}{}^\dagger \tilde \gamma_4\;
 \hat \gamma_5^{(-n)} \; \tilde \gamma_4 \tilde U^{(n)} \tilde \gamma_4 =
\gamma_5, \label{plus_to_minus}
\end{eqnarray}
and therefore
\begin{eqnarray}
\tilde U^{(-n)} = \tilde \gamma_4 \tilde U^{(n)} \tilde
\gamma_4. \label{U_to_minus_U}
\end{eqnarray}
Equation (\ref{U_to_minus_U}) need not be valid for $n=0$
(there is, in fact, a degeneracy) and this case has to be
considered separately, which will be done in Section~\ref{CPTnoninvariance}.

\subsection{CPT invariance in the $n \neq 0$ sectors}\label{CPTinvariance}
In this subsection, we first
examine the CPT transformation of the bases $v_j^{(n)} (\vec x ;U)$
for the $n \neq 0$ Fourier modes (assuming odd $N \geq 3$).
Concretely, we have to find the transformation behavior of
the operator $\tilde U^{(n)}[U]$ under CPT; cf. Eq.~(\ref{vjn}).
This is relatively straightforward and the reader who is not interested in the
technical details may skip ahead to the result (\ref{CPTNichtNull}).

The behavior of the operator $\hat \gamma_5^{(n)} \equiv \gamma_5 V^{(n)}$
under a CPT transformation follows from Eqs.~(\ref{cptV}) and (\ref{Vdagger}):
\begin{eqnarray}\label{gamma5hatnCPT}
\mathcal{R}\, \hat \gamma_5^{(n)} [U]\, \mathcal{R} = \gamma_5\, \hat
\gamma_5^{(-n)} [U^\theta]\, \gamma_5.
\end{eqnarray}
 From the definition (\ref{gamma5hatn}) and the relation (\ref{prop1}), we have
\begin{eqnarray}
\hat \gamma_5^{(-n)} [U] =
\left(
\begin{array}{cc}
-h_3^{(n)}[U]   &\; ih_2^{(n)}{}^\dagger [U] \\
-ih_2^{(n)} [U] &\; -h_1^{(n)} [U]
\end{array}
\right),
\end{eqnarray}
so that the CPT-transformed of $\hat \gamma_5^{(n)}$ is
\begin{eqnarray}
\mathcal{R}\, \hat \gamma_5^{(n)} [U]\, \mathcal{R} = \left(
 \begin{array}{cc}
-h_3^{(n)}[U^\theta]  &\;\; -ih_2^{(n)}{}^\dagger [U^\theta] \\
ih_2^{(n)} [U^\theta] &\;\; -h_1^{(n)} [U^\theta]
\end{array}
\right). \label{cpt_gamma5}
\end{eqnarray}

The operator $\tilde U^{(n)}[U]$ can be written as follows:
\begin{eqnarray}\label{Untilde}
\tilde U^{(n)}[U] = \left(
\begin{array}{cc}
\alpha^{(n)} [U] &\;\; \beta^{(n)} [U] \\[0.2cm]
\gamma^{(n)} [U] &\;\; \delta^{(n)} [U]
\end{array}
\right),
\end{eqnarray}
with $2\times 2$ spinor matrices $\alpha^{(n)}$, $\beta^{(n)}$,
$\gamma^{(n)}$ and $\delta^{(n)}$.
According to Eq.~(\ref{diag}),
these matrices have to satisfy the following relations:
\begin{eqnarray}
\left(\1 - h_1^{(n)}\right) \alpha^{(n)} &=& ih_2^{(n)} \gamma^{(n)},
\label{bed1}\\
\left(\1 + h_3^{(n)}\right) \delta^{(n)} &=& ih_2^{(n)}{}^\dagger
\beta^{(n)}. \label{bed2}
\end{eqnarray}
The CPT transform of the operator $\tilde U^{(n)}$ is
\begin{eqnarray}\label{UntildeCPT}
\gamma_5 \mathcal{R}\: \tilde U^{(n)} [U]\, \mathcal{R} \gamma_5
= \left(
\begin{array}{cc}
\mathcal{R} \alpha^{(n)} [U]\mathcal{R} &\;\; -\mathcal{R}\beta^{(n)} [U]
\mathcal{R} \\
-\mathcal{R}\gamma^{(n)} [U]\mathcal{R} &\;\; \mathcal{R}\delta^{(n)}
[U]\mathcal{R}
\end{array}
\right).
\end{eqnarray}
Multiplying the relations (\ref{bed1}), (\ref{bed2}) with the
coordinate reflection operator $\mathcal{R}$ from the right and the left,
and using Eq.~(\ref{cpt_gamma5}), gives
\begin{eqnarray}
\left(\1 + h_3^{(n)}\right) \mathcal{R} \alpha^{(n)} \mathcal{R} &=&
-ih_2^{(n)}{}^\dagger \mathcal{R} \gamma^{(n)} \mathcal{R}, \nn\\
\left(\1 - h_1^{(n)}\right) \mathcal{R} \delta^{(n)} \mathcal{R} &=&
-ih_2^{(n)} \mathcal{R}\beta^{(n)}\mathcal{R}.
\end{eqnarray}
Hence, the block matrices  transform as follows:
\begin{eqnarray}
\!\!\!\mathcal{R} \alpha^{(n)} [U] \mathcal{R} &=& \delta^{(n)} [U^\theta] R_1^{(n)}
[U^\theta] , \;\;
\mathcal{R} \gamma^{(n)}[U]  \mathcal{R} = - \beta^{(n)}
[U^\theta] R_1^{(n)} [U^\theta], \nonumber \\[0.2cm]
\!\!\mathcal{R} \delta^{(n)} [U] \mathcal{R} &=& \alpha^{(n)} [U^\theta] R_2^{(n)}
[U^\theta] , \;\;
\mathcal{R} \beta^{(n)}[U]  \mathcal{R} = - \gamma^{(n)}
[U^\theta] R_2^{(n)} [U^\theta],
\end{eqnarray}
with unitary $2 \times 2$ matrices $R_a^{(n)}$ which depend on $\alpha^{(n)},
\ldots, \delta^{(n)}$. The matrices $R_1^{(n)}$ and $R_2^{(n)}$ are, however,
not independent. Multiplying (\ref{bed1}) from the left and right with $\1 =
\mathcal{R} \mathcal{R}$, one gets
\begin{eqnarray}
R_2^{(n)} [U] \mathcal{R} R_1^{(n)} [U^\theta] \mathcal{R} &=& \1.
\end{eqnarray}
Similarly,  one gets from Eq.~(\ref{bed2})
\begin{eqnarray}
R_1^{(n)} [U] \mathcal{R} R_2^{(n)} [U^\theta] \mathcal{R} &=& \1.
\end{eqnarray}

We can now choose the matrix (\ref{ambig}) to be given by
\begin{eqnarray}
Y^{(n)} = \left(
\begin{array}{cc}
R_2^{(n)} [U] &\;\; 0 \\
0             &\;\; \1
\end{array}
\right).
\end{eqnarray}
For this particular choice of $Y^{(n)}$, the  $2\times 2$ spinorial matrices in
$\tilde U^{(n)\,\prime} = \tilde U^{(n)}\,Y^{(n)}$ are given by
\begin{eqnarray}
\alpha^{(n)}{}'[U] &=& \alpha^{(n)}[U] R_2[U] , \qquad \delta^{(n)}{}'[U] =
\delta^{(n)}[U], \nonumber\\[0.2cm]
\gamma^{(n)}{}'[U] &=& \gamma^{(n)}[U] R_2[U] , \qquad \beta^{(n)}{}'[U] =
\beta^{(n)}[U],
\end{eqnarray}
and the CPT-transformed matrices are simply
\begin{eqnarray}\label{alphaprimeCPT}
\mathcal{R}\, \alpha^{(n)}{}' [U]\, \mathcal{R} &=& \delta^{(n)}{}' [U^\theta],
\qquad
\mathcal{R}\, \gamma^{(n)}{}'[U]\, \mathcal{R} = - \beta^{(n)}{}'
[U^\theta], \nonumber\\[0.2cm]
\mathcal{R}\, \delta^{(n)}{}' [U]\, \mathcal{R} &=& \alpha^{(n)}{}' [U^\theta],
\qquad
\mathcal{R}\, \beta^{(n)}{}'[U]\, \mathcal{R} = - \gamma^{(n)}{}' [U^\theta].
\end{eqnarray}
Henceforth, we drop the primes.

 From Eqs.~(\ref{Untilde}), (\ref{UntildeCPT}), and (\ref{alphaprimeCPT}),
without the primes, the needed CPT transform of the operator $\tilde U^{(n)}[U]$
follows,
\begin{eqnarray}
\gamma_5 \mathcal{R}\: \tilde U^{(n)} [U]\, \mathcal{R} \gamma_5 = \tilde \gamma_4
\,\tilde U^{(n)} [U^\theta]\, \tilde \gamma_4. \label{CPTNichtNull}
\end{eqnarray}
With this result, the change of the measure can be calculated.
The relevant transformation matrix in Eq.~(\ref{Qdefinition}) reads
\begin{eqnarray}\label{Qminncalculation}
\mathcal{Q}^{(-n)}_{kl} [U] &=& a^3 \sum_{\vec x}
  v_k^{(-n)}{}^\dagger (\vec x; U)\, \gamma_5 \mathcal{R}\, v_l^{(n)} (\vec
  x; U^\theta)\nonumber \\
&=&
  a^3 \sum_{\vec x}\left( \varphi_k^\dagger(\vec x)\, \tilde
  U^{(-n)}{}^\dagger [U] \right) \mathcal{R} \gamma_5 \left(\tilde U^{(n)}
  [U^\theta]\, \varphi_l(\vec x) \right)\nonumber \\
&=&
  a^3 \sum_{\vec x} \varphi_k^\dagger(\vec x)\, \tilde U^{(-n)}{}^\dagger[U]\;
   \tilde \gamma_4 \tilde U^{(n)} [U] \tilde \gamma_4\;
  \mathcal{R} \gamma_5\, \varphi_l(\vec x).
\end{eqnarray}
For $n \not= 0$, the identity (\ref{U_to_minus_U}) then gives
\begin{eqnarray}\label{Qminn}
\mathcal{Q}^{(-n)}_{kl} [U] = a^3 \sum_{\vec x}
  \varphi_k^\dagger(\vec x)\, \mathcal{R} \gamma_5 \varphi_l(\vec x),
\end{eqnarray}
which is independent of the link variables $U_m (\vec x)$. The matrices
$\bar{\mathcal{Q}}^{(n)}$ are also independent of the link variables.
Therefore, only $\det \mathcal{Q}^{(0)}$ can give a
change of the fermion integration measure which depends on the link variables.
But before we turn to the $n=0$ contribution in the next
subsection, there is one technical point that needs to be clarified.

The transformation matrices $\mathcal{Q}^{(n)}$ and $\bar{\mathcal{Q}}^{(n)}$
for $n \not= 0$  are nonlocal expressions,
due to the presence of the reflection operator $\mathcal{R}$ in Eq.~(\ref{Qminn})
and a similar equation for  $\bar{\mathcal{Q}}^{(n)}$.
But these matrices have to be combined, since
we are only interested in the simultaneous
change of the measures $\prod_j \d c_j^{(n)}$ and $\prod_k \d \bar
c_k^{(n)}$; see  Eqs.~(\ref{integral}) and (\ref{GammaUtheta}).
The combined transformation matrix for a Fourier mode $n\neq 0$ is, in fact,
\begin{eqnarray}
\sum_l \mathcal{Q}^{(n)}_{kl} \bar{\mathcal{Q}}^{(n)}_{lm}
 &=& \sum_l \left(a^3 \sum_{\vec x} \varphi_k^\dagger (\vec x)
 \mathcal{R} \gamma_5 \varphi_l(\vec x) \right) \left(a^3 \sum_{\vec y}
 \bar \varphi_l (\vec y) \mathcal{R} \gamma_5 \bar
 \varphi_m^\dagger(\vec y) \right)\nonumber\\
&=& a^3 \sum_{\vec x} \big( 0, \psi_k^\dagger(\vec x) \big)\,
 \mathcal{R} \gamma_5
\left(
\begin{array}{cc}
0  &\;\; 0 \\
\1 &\;\; 0 \end{array}
\right)
\gamma_5 \mathcal{R} \left( \begin{array}{c}
\psi_m (\vec x) \\ 0
\end{array}\right) \nonumber \\
&=& -a^3 \sum_{\vec x} \big( \psi_k^\dagger(\vec x), 0 \big)
\left( \begin{array}{c}
\psi_m (\vec x) \\ 0 \end{array}\right) =
-\delta_{km}. \label{QnQnbar}
\end{eqnarray}
For the second equality in Eq.~(\ref{QnQnbar}),
the explicit form of the bases (\ref{phibar}) and (\ref{phi}), with
$\bar \psi_k(\vec x)$ $=$ $\psi_k^\dagger(\vec x)$, has been used,
together with the completeness of the two-spinor basis $\psi_k(\vec x)$,
\begin{eqnarray}
\sum_l \psi_l (\vec x)\: \psi^\dagger_l(\vec y) = a^{-3}\, \1 \;
\delta_{\vec x \vec y},
\end{eqnarray}
where $\1$ is the unit operator in spinor space.

The determinant of the matrix (\ref{QnQnbar}) is 1, because there is an
even number of eigenvalues $-1$.
Together with Eqs.~(\ref{integral}) and (\ref{GammaUtheta}),
this shows the CPT invariance of the effective action in the $n \neq 0$
sectors, at least for the case of odd $N$.

\subsection{CPT noninvariance in the $n=0$ sector}\label{CPTnoninvariance}
We now turn to the contribution of the $n=0$ Fourier mode to
the change of the effective action under CPT
(for odd $N$, the other Fourier modes do not contribute, as we have seen).
The $n=0$ sector produces an essentially massless three-dimensional fermion;
cf. Section~\ref{Effe}. It is, of
course, known that the effective action of a $U(1)$ gauge theory with a
single massless Dirac fermion in three dimension displays the so-called
``parity'' anomaly \cite{R84,ADM85,CL89,KN98,BN01}. We can,
therefore, expect to find an anomaly also in our case.

Using the results (\ref{gamma5hatnCPT})--(\ref{Qminncalculation}),
which hold also for
$n=0$, the transformation matrix $\mathcal{Q}^{(0)}[U]$ is
\begin{eqnarray} \label{Q0}
\mathcal{Q}^{(0)}_{kl}[U] =
a^3 \sum_{\vec x} \varphi_k^\dagger(\vec x)\,
\tilde U^{(0)}{}^\dagger [U]\; \tilde \gamma_4 \tilde U^{(0)} [U] \tilde
\gamma_4\; \mathcal{R} \gamma_5\, \varphi_l(\vec x).
\end{eqnarray}
Now, an explicit solution for the operator $\tilde U^{(0)}$ from
Eq.~(\ref{diag}) can be found:
\begin{eqnarray}
\tilde U^{(0)} = \half \left(
\begin{array}{cc}
\1 + W^{(0)} &\;\; -(\1 - W^{(0)}{}^\dagger) \\
\1 - W^{(0)} &\;\; \1 + W^{(0)}{}^\dagger
\end{array}
\right), \label{Utilde0}
\end{eqnarray}
with the unitary operator
\begin{eqnarray}\label{W0}
W^{(0)} \equiv h_1^{(0)} + i h_2^{(0)},
\end{eqnarray}
in terms of the entries of  $\hat \gamma_5^{(0)}$
as defined by Eq.~(\ref{gamma5hatn}).
The validity of the solution (\ref{Utilde0}) rests on the following results.
First, the relation (\ref{prop1}) holds also
for $n=0$, so that $h_3^{(0)} = - h_1^{(0)}$ and
$h_2^{(0)}= h_2^{(0)\,\dagger}$. Second, the unitarity of
$\hat \gamma_5^{(0)}$ implies that $(h_1^{(0)})^2 +
(h_2^{(0)})^2 = \1$ and that $h_1^{(0)}$ commutes with $h_2^{(0)}$.
Third, this makes for a unitary operator $W^{(0)}$.
(These results, including Eq.~(\ref{Utilde0}), are essentially
contained in Ref.~\cite{BN01}.)

The transformation matrix (\ref{Q0}) is then
\begin{eqnarray}
\mathcal{Q}^{(0)}_{kl} [U]= - a^3 \sum_{\vec x} \psi^\dagger_k(\vec x)
\,W^{(0)}[U]\, \psi_l(-\vec x),
\end{eqnarray}
for the explicit form (\ref{phi}) of the $n=0$ basis.
The result for $\mathcal{Q}^{(0)}[U]$ is again nonlocal, but there is
also the contribution of the matrix $\bar{\mathcal{Q}}^{(0)}$. The
product of these two matrices is local,
\begin{eqnarray}\label{Q0Q0bar}
\sum_l \mathcal{Q}^{(0)}_{kl}[U]\, \bar{\mathcal{Q}}^{(0)}_{lm} = - a^3
\sum_{\vec x} \psi^\dagger_k(\vec x)\, W^{(0)}[U]\, \psi_m(\vec x),
\end{eqnarray}
just as was the case in the previous subsection.

 From Eqs.~(\ref{GammaUtheta}), (\ref{QnQnbar}) and (\ref{Q0Q0bar}),
the total change of the effective action (\ref{integral})
under a CPT transformation (\ref{Utheta}) of the gauge field configuration
considered is then given by
\begin{eqnarray}\label{DeltaGamma}
\Delta \Gamma[U] \equiv \Gamma[U^\theta] - \Gamma[U] = - \ln \det \left(-
  a^3 \sum_{\vec x} \psi^\dagger_k(\vec x)\, W^{(0)}[U]\, \psi_m(\vec x)\right),
\end{eqnarray}
which is purely imaginary because of the unitarity of $W^{(0)}[U]$.
It is not difficult to show that the \rhs~of Eq.~(\ref{DeltaGamma}) vanishes
for $U_m(\vec x)=1$.

The result (\ref{DeltaGamma}) makes clear that, for the $U(1)$ gauge field
configurations (\ref{config})
and odd $N$, the CPT anomaly resides entirely in the $n=0$ factor of
Eq.~(\ref{integral}). This means that the CPT anomaly for the gauge field
configurations  considered effectively
reduces to the three-dimensional ``parity'' anomaly \cite{R84,ADM85} mentioned
above, which has also been studied on the lattice \cite{CL89,KN98,BN01}. In
fact, Ref.~\cite{KN98} has reported some numerical results for
the determinant of Eq.~(\ref{DeltaGamma}). These numerical results
show that the determinant (\ref{DeltaGamma})
differs from 1 for appropriate three-dimensional lattice gauge field
configurations.\footnote{The three-dimensional $U(1)$ gauge field
configuration considered in Section 4 of Ref.~\cite{KN98} has flux
quantum numbers $m_{12}\neq 0$ and $m_{13}=m_{23}=0$. The corresponding
four-dimensional configuration (\ref{config}) has also
$m_{14}=m_{24}=m_{34}=0$, so that there is no fermion number violation
and the matrices $\mathcal{Q}^{(0)}$ and $\bar{\mathcal{Q}}^{(0)}$
have equal dimensions; cf. the sentence below Eq.~(\ref{Qdefinition}).}
An analytical result will be given in Section \ref{Ntwo}.

At this point, the question arises as to how the apparently well-defined
chiral lattice gauge theory of Ref.~\cite{L99} could have produced a violation
of CPT invariance. The crucial observation is that the particular solution
(\ref{Utilde0}) for the gauge-covariant
operator $\tilde U^{(0)}[U]$, which determines the
$n=0$ part of the spinor bases (\ref{vjn}), is not invariant under a CPT
transformation.
For generic gauge field configurations (\ref{config}), one has, in fact,
\begin{eqnarray}\label{CPTcrux}
\gamma_5 \mathcal{R}\, \tilde U^{(0)} [U]\, \mathcal{R} \gamma_5 =
\tilde \gamma_4\, \tilde U^{(0)}[U^\theta]\, \tilde \gamma_4 =
\tilde U^{(0)} [U^\theta]\, \tilde B [U^\theta]
\not= \tilde U^{(0)} [U^\theta],
\end{eqnarray}
with the block-diagonal matrix
\begin{eqnarray}
\tilde B[U^\theta] = \left(
\begin{array}{cc}
W^{(0)\dagger}[U^\theta] & 0 \\
0 & W^{(0)}[U^\theta]
\end{array}
\right),
\end{eqnarray}
where $W^{(0)}$ and $W^{(0)\dagger}$ act on left- and right-handed fermions,
respectively. Hence, the CPT transformation leads in general
to another theory, with different basis spinors.
This new theory can be transformed back to the old one,
but there appears a Jacobian, $\det W^{(0)}$.
The CPT anomaly occurs if this determinant is different from
unity.\footnote{As emphasized in Ref.~\cite{L00},
the fermion integration measure is not a product of local measures.
The CPT anomaly appears precisely in the $n=0$ sector, which corresponds to
spinor fields that are constant in the 4 direction.
We intend to elaborate on the issue of locality and gauge invariance
in a forthcoming publication.}

For completeness, we remark that the theory defined with the operator
$\tilde U^{(0)}$ from Eq.~(\ref{Utilde0}) replaced by
$\tilde \gamma_4 \tilde U^{(0)} \tilde \gamma_4$ has the same
$\Delta \Gamma[U]$, but with a different sign.
More generally, there is a freedom in the choice of $\tilde U^{(0)}$,
which leads to an overall factor $(2\, k_0+1)$, for $k_0 \in \Z$, on the
\rhs~of Eq.~(\ref{DeltaGamma}).
Further details are given in Appendix~\ref{Fermionicmeasure};
see also Section 3 of Ref.~\cite{CL89} for a general discussion of the
factor $(2\, k_0+1)$ in the three-dimensional ``parity'' anomaly.
For the rest of this paper, we keep the theory with the $n=0$ spinor
basis defined by the operator $\tilde U^{(0)}$ from Eq.~(\ref{Utilde0}).

In conclusion, the \rhs~of Eq.~(\ref{DeltaGamma}) is nonzero for generic
gauge field configurations of the type (\ref{config}),
which establishes the existence of a CPT anomaly in the theory considered.
The CPT anomaly occurs because of the noninvariance
of the measure in the fermionic integral (\ref{integral}); cf. Ref.~\cite{F80}.

\section{CPT anomaly for even $N$}\label{evenN}
The results of the previous section also hold for even $N$,
except for an additional contribution of the $n = N/2$ Fourier mode;
cf. Eqs.~(\ref{nodd})--(\ref{neven}).
For this case, Eq.~(\ref{prop1}) is replaced by
\begin{eqnarray}
\hat \gamma_5^{(N/2)} \tilde \gamma_4 = - \tilde \gamma_4 \hat
\gamma_5^{(N/2)},
\end{eqnarray}
and the consequences of this property need to be investigated. In fact,
we can just repeat the $n=0$ calculation of Section \ref{CPTnoninvariance},
with the replacement $h^{(0)}_a \to h^{(N/2)}_a$.
This gives an additional transformation matrix
\begin{eqnarray}
\sum_l \mathcal{Q}^{(N/2)}_{kl}\, \bar{\mathcal{Q}}^{(N/2)}_{lm} = - a^3
\sum_{\vec x} \psi^\dagger_k(\vec x)\, W^{(N/2)}\, \psi_m(\vec x),
\end{eqnarray}
with the unitary operator
\begin{eqnarray}
W^{(N/2)} \equiv h_1^{(N/2)} + i h_2^{(N/2)}.
\end{eqnarray}
Hence, the total change of the measure is
\begin{eqnarray}
&&\det \left( \sum_l \mathcal{Q}^{(0)}_{kl}[U]\,
                     \bar{\mathcal{Q}}^{(0)}_{lm}\right)\;
    \det \left( \sum_l \mathcal{Q}^{(N/2)}_{kl}[U]\,
                        \bar{\mathcal{Q}}^{(N/2)}_{lm}\right)
    =\nonumber \\
&&\det \left(- a^3 \sum_{\vec x} \psi^\dagger_k(\vec x)\, W^{(0)}[U]\,
  \psi_m(\vec x)\right)\nonumber\\
&& \times \det \left(- a^3 \sum_{\vec x}
  \psi^\dagger_k(\vec x)\, W^{(N/2)}[U]\, \psi_m(\vec
    x)\right). \label{changeEvenN}
\end{eqnarray}
The first determinant on the right hand side of Eq.~(\ref{changeEvenN})
is the same as found for odd $N$ and second determinant
approaches 1 in the continuum limit $a \to 0$
(for a given smooth gauge field configuration), as will become clear in
the next section.

\section{Effective action for $N = 2$} \label{Ntwo}
The result (\ref{changeEvenN}) and the basic input (\ref{Xn}), evaluated
for $n=N/2$,  make clear that
the CPT anomaly for even $N$ is independent of the size $N$ (but not of $a$).
We can, therefore, restrict the calculations to the
case of $N=2$.

A straightforward calculation, starting from Section~\ref{Effe},
gives a relatively simple result for the $N=2$ effective action\footnote{The
crucial simplification for $N=2$ occurs in Eq.~(\ref{Xn}),
with the $\gamma_4$ term dropping out. The spinor fields can then be transformed
with an appropriate constant unitary matrix.
The resulting operators (\ref{3Dop})
have already appeared in the context of the overlap formalism in three
dimensions; see Section 2.2 of Ref.~\cite{KN98}.}:
\begin{eqnarray}
\!\!\!\!\!
\exp\left(-\Gamma^{(N=2)}[U]\right) &=&
\tilde K \,\prod_{n=0}^1 \int \prod_l \d c^{(n)}_l \prod_k \d \bar c^{(n)}_k
\exp \left(- \sum_{k,l} \bar c_k^{(n)}\, \tilde M_{kl}^{(n)}[U]\,
                             c_l^{(n)}\right),\nn\\
&&
\end{eqnarray}
with the matrices
\begin{eqnarray} \label{Mtilde}
\tilde M_{kl}^{(n)}[U] = a^3 \sum_{\vec x} \psi_k^\dagger (\vec x)\,
\big( \1- \tilde V^{(n)}[U] \big)\, \psi_l(\vec x),
\end{eqnarray}
and the unitary operators
\begin{eqnarray}
\tilde V^{(n)} \equiv \tilde X^{(n)} \left(\tilde X^{(n)\,\dagger} \tilde
  X^{(n)} \right)^{-1/2}, \label{3Dop}
\end{eqnarray}
for
\begin{eqnarray}
\tilde X^{(0)} &\equiv& \1 - a\left( \half \sum_{\mu=1}^3 \left( \sigma_\mu
  (\nabla_\mu + \nabla_\mu^*) - a\nabla_\mu^*\nabla_\mu\right) \right),\nn \\
\tilde X^{(1)} &\equiv& \1 - a\left( \half \sum_{\mu=1}^3\left( \sigma_\mu
  (\nabla_\mu + \nabla_\mu^*) - a\nabla_\mu^*\nabla_\mu\right) +
  2/a \right) .
\end{eqnarray}

In the classical continuum limit $a \to 0$ (with fixed $L' = N'\, a$),
one replaces the inverse square root of Eq.~(\ref{3Dop}) by the identity operator.
For $n=0$, one essentially
obtains in Eq.~(\ref{Mtilde}) the three-dimensional Wilson-Dirac operator with
a positive mass of order $a/L^{\prime\, 2}$, whereas
for $n=1$ one has the three-dimensional Wilson-Dirac operator with a
negative mass $m= -2/a$. The effective actions for these operators have been
calculated in Ref. \cite{CL89} to second order in the (bare) coupling
constant $e$. An anomalous term is present for $n=0$ but not for $n=1$
(in our case, the Wilson parameter $s$ of Ref.~\cite{CL89} equals $-1$).

For $A_4 =0$ and smooth continuum gauge fields $A_m(\vec x)$ of small amplitude
(i.e., perturbative gauge fields), the result
for the imaginary part of the effective action is
\begin{eqnarray}
\mathrm{Im}\,\Gamma^{(N=2)} [A\,] \approx e^2 \;(2\pi + 0)\; \Omega_{\rm
  CS}[A\,] ,\label{effactN2}
\end{eqnarray}
in terms of the \CS~integral
\begin{eqnarray}
\Omega_{\rm CS}[A\,]  \equiv (16\, \pi^2)^{-1}
\int_{V'} \d^3 x \; \epsilon^{klm} A_k(\vec x)\, \partial_l A_m(\vec x),
\end{eqnarray}
with $\epsilon^{klm}$ the completely antisymmetric Levi-Civita symbol,
normalized to $\epsilon^{123}=1$.
As mentioned above, the factor $2\pi$ in Eq.~(\ref{effactN2}) traces
back to the $n=0$ Fourier mode of the
single fermion considered and the factor $0$ to the $n=1$ mode.

The effective action (\ref{effactN2}) is obviously noninvariant
under the CPT transformation
$A_m (\vec x) \to - A_m (-\vec x)$;
cf. Eqs.~(\ref{reflection}) and (\ref{Utheta}).
This shows that the first determinant on
the right hand side of Eq.~(\ref{changeEvenN}),
which is precisely the one of Eq.~(\ref{DeltaGamma}), differs from 1, at
least for appropriate gauge fields in the continuum limit.\footnote{The
result (\ref{effactN2}) shows that the CPT violation found here traces back to
the Wilson term $a\nabla^2$ in Eq.~(\ref{DW3dim}), together with the antiperiodic
\bcs~(\ref{apbc}). A similar role was played by the Pauli--Villars mass
in the original calculation of Ref.~\cite{K00}.}
Since the $n= N/2 = 1$ mode does not contribute at all to
Eq.~(\ref{effactN2}), it is also clear that the second determinant on
the right hand side of Eq.~(\ref{changeEvenN}) approaches 1 in the
continuum limit.

Because the gauge fields at $x_4 = 0$ and $x_4 = a$ are identical, the
result (\ref{effactN2}) can trivially be written as follows ($L=N\,a =2\,a$):
\begin{eqnarray}
\mathrm{Im}\,\Gamma^{(N=2)} [A\,] &\approx&
e^2\; \half \sum_{n_4=0}^1 \; 2\pi\,
\Omega_{\rm CS} \left[ \vec A(\vec x, n_4\, a)\right] \nonumber\\
&=&
e^2\;  a \sum_{n_4=0}^1\;  \frac {2\pi}L\,
\Omega_{\rm CS} \left[ \vec A(\vec  x, n_4\, a) \right],
\label{cont_eff_act2}
\end{eqnarray}
which is of the same form as the \CS-like term (4.1) of Ref.~\cite{K00}.
Here, the circle in the 4 direction is replaced by $N=2$ links,
which effectively close because of the periodic \bcs~(\ref{pbcs})
of the gauge fields.

Up till now, we have considered $x_4$-independent gauge
fields (\ref{config}). But our calculation also holds for the
following gauge field configurations:
\begin{eqnarray}
U^\prime_4 (\vec x, n_4\, a) &=&
\exp\big(   i \omega(\vec x, n_4\, a)
          - i \omega(\vec x, n_4\, a + a)\big),
\nn \\[0.2cm]
U^\prime_m (\vec x, n_4\, a) &=&
\exp\big(  i \omega(\vec x,           n_4\, a)\big)\,U_m (\vec x)\,
\exp\big(- i \omega(\vec x +a\hat{m}, n_4\, a)\big),
\label{confignew}
\end{eqnarray}
with an arbitrary periodic function $\omega(x)$.
[The four-dimensional gauge field configuration (\ref{confignew})
is, of course, a gauge transform of the configuration (\ref{config}).]
At this point, we need to return to
the original chiral $U(1)$ gauge theory (\ref{theory})
of $N_F=9$ left-handed fermions with charges
$q_f=+1$, for $f=1, \cdots, 8$, and $q_9=-2$.

For the case of $N=2$ links in the 4 direction, the imaginary part of the
effective action is then
\begin{eqnarray}
\mathrm{Im}\,\Gamma^{(N=2)} [A'\,] = \mathrm{Im}\,\Gamma^{(N=2)} [A\,]
\approx
12\,e^2\;  a \sum_{n_4=0}^1\;  \frac {2\pi}L\,
\Omega_{\rm CS} \left[ \vec A^{\,\prime}(\vec  x, n_4\, a) \right],
\label{cont_eff_act2Aprime}
\end{eqnarray}
with the $x_4$-dependent perturbative gauge fields
$A^\prime_m(\vec  x, n_4\, a)$ from Eq.~(\ref{confignew})
and the factor $e^2$ from Eq.~(\ref{cont_eff_act2}) replaced by
$\sum (q_f\,e)^2$. The result found is essentially the same as before, because
the factors $\exp(\pm i\omega)$ in Eq.~(\ref{confignew})
can be absorbed by a change of variables of the
fermion fields. The corresponding Jacobian is 1, as long as the
gauge anomaly cancels between the different fermion
species. It has been shown in Ref.~ \cite{L99} that this anomaly cancellation
occurs provided $\sum (q_f)^3=0$, which is the
same condition as for the continuum theory \cite{A69,BJ69,BIM72,GJ72}.

In conclusion, we have calculated (using the results of
Ref.~\cite{CL89}) the effective action for the case of $N=2$ links in
the periodic direction and for perturbative gauge field configurations
(\ref{confignew}). This effective action, Eq.~(\ref{cont_eff_act2Aprime}),
represents a first result for a genuinely four-dimensional chiral lattice
gauge theory, albeit of very small periodic dimension ($L= 2\, a \to
0$).\footnote{The ``trivial'' case of a single link in the 4 direction
    ($N=1$) has already been considered in the overlap formalism \cite{KN98}.}

\section{Discussion} \label{Discussion}

In this paper, we have rigorously established the existence of a CPT anomaly
in a particular lattice version of the  Abelian chiral gauge theory (\ref{theory}).
This was done for the class of $x_4$-independent gauge fields
with trivial holonomies; see Eqs.~(\ref{config}) and (\ref{holonomy}).
[The periodic compact dimension has coordinate $x_4$ and size $L = N a$,
with lattice spacing $a$.]
For an arbitrary odd number $N$ of links in the
periodic direction, the noninvariance of the effective gauge
field action under a CPT transformation holds for arbitrary values
of the lattice spacing $a$. In other words, the result
(\ref{DeltaGamma}) is valid nonperturbatively.

Having established the noninvariance of the effective action under
CPT, the question arises as to what the imaginary part of the
effective action really is. The full answer is, of course, not
known. But we have obtained a partial result, namely for $N=2$
links in the periodic direction and in the continuum limit $a \to 0$
(the smooth perturbative gauge field being held fixed). The
result, Eq.~(\ref{cont_eff_act2Aprime}), agrees with the \CS-like term
of Ref. \cite{K00}. It is hoped that further results can be obtained in this way.

Finally, we would like to raise another question. Namely, does
the Euclidean chiral lattice gauge theory considered, i.e., the
theory given by Eq.(\ref{theory}), have an anomalous violation of
reflection positivity in the periodic $x_4$ coordinate? (See, e.g.,
Ref.~\cite{MM94} for further details on reflection positivity and its
relation to unitarity in the physical theory  with a Lorentzian signature of
the metric.) The question
arises, because the \CS-like term of the Euclidean continuum theory is known
to violate $x_4$ reflection positivity \cite{AK01}.
In fact, this is precisely the reason for demanding that the
periodic compact dimension ``responsible'' for the CPT anomaly be spacelike
in the theory with a Lorentzian metric. Clearly, the issue of reflection
positivity (unitarity) in chiral lattice gauge theories like Eq.~(\ref{theory})
deserves further study.

\section*{Note added} \label{Noteadded}
After the completion of this work, we have become aware of
Ref.~\cite{FIS02}, which discusses the explicit and anomalous
breaking of CP in \chlgth. Without Higgs field, the CP breaking is
expected to disappear for a suitable continuum limit. This behavior
would differ from the CPT anomaly found here, which persists in the continuum
limit; cf. Eqs.~(\ref{DeltaGamma}) and (\ref{cont_eff_act2Aprime}).
Still, the \emph{inherent} breaking of CP and T
in \chlgth~(see Ref.~\cite{FIS02} and references
therein) may be directly relevant to the fundamental origin of the
CPT anomaly \cite{K00,K01}.

%%\begin{note} xxx \end{note}   %%does not work
%%\begin{ack} xxx \end{ack}     %%works

\begin{appendix}
\section{Notation}\label{Notation}
The $U(1)$ lattice gauge theory considered has already been presented in
Section \ref{general} of the main text.
In this appendix, we briefly review some further notation, which
basically follows Ref.~\cite{L99}.

The gauge-covariant forward- and backward-difference operators are defined by
\begin{eqnarray}
\nabla_\mu \psi(x) &\equiv& \big(\,
                                 R[U_\mu(x)]\, \psi(x+a \hat{\mu}) - \psi(x)\,
                            \big)/a,
\\[0.2cm]
\nabla_\mu^* \psi(x) &\equiv& \big(\,\psi(x) -
        R[U_\mu^\dagger(x-a \hat{\mu})]\,\psi(x-a \hat{\mu})
                               \big)/a,
\end{eqnarray}
with the lattice spacing $a$ and the unit vector  $\hat \mu$ in the
$\mu$ direction. Here, $R[U]$ indicates the unitary representation of the
fermion considered. For an Abelian $U(1)$ gauge theory,  $R[U]$ is
simply $U^{\,q}$, with $q$ the integer charge of the fermion and $U$ the link
variable written as Eq.~(\ref{expieaA}) in the main text. For the theory
(\ref{theory}), for example,
the left-handed fermions have integer charges $q=1$ and $q=-2$.

Throughout this paper, we use the following representation of the Dirac
matrices:
\begin{eqnarray}\label{gammamu}
\gamma_\mu = \left(
\begin{array}{cc}
0                  &\; \sigma_\mu \\
\sigma_\mu^\dagger &\; 0
\end{array}
\right),
\end{eqnarray}
in terms of the three Pauli matrices
\begin{eqnarray}
\sigma_1 = \left(
\begin{array}{cc}
0 &\;\; 1 \\
1 &\;\; 0
\end{array}
\right) , \quad
\sigma_2 = \left(
\begin{array}{cc}
0 &\; -i \\
i &\; 0
\end{array}
\right), \quad
\sigma_3 = \left(
\begin{array}{cc}
1 &\; 0 \\
0 &\; -1
\end{array}
\right),
\end{eqnarray}
and
\begin{eqnarray}
\sigma_4 = i\, \1 \equiv
\left(
\begin{array}{cc}
i &\;\; 0 \\
0 &\;\; i
\end{array}
\right).
\end{eqnarray}
The corresponding chirality matrix
$\gamma_5 \equiv \gamma_1 \gamma_2 \gamma_3 \gamma_4$ is diagonal,
\begin{eqnarray}
\gamma_5 = \left(
\begin{array}{cc}
\1 &\; 0 \\
0  &\; -\1
\end{array}
\right).
\end{eqnarray}

\section{Fermion integration measure in the $n=0$ sector}\label{Fermionicmeasure}
In Section~\ref{CPTnoninvariance}, we have given an explicit example of the
operator $\tilde U^{(0)}$,
which determines the  spinor basis  in the $n=0$ sector,
according to Eq.~(\ref{vjn}). But $\tilde U^{(0)}$ as given
by Eq.~(\ref{Utilde0}) and the corresponding alternative
$\tilde \gamma_4 \tilde U^{(0)} \tilde \gamma_4$ are not
the only possibilities (up to constant matrices, of course).
This can be shown as follows.

As mentioned above Eq.~(\ref{ambig}), the operator $\tilde U^{(0)}$
is not unique and we can, for example, replace the $\tilde U^{(0)}$
of Eq.~(\ref{Utilde0}) by
\begin{eqnarray}
\tilde U^{(0)}{}' = \tilde U^{(0)} \left(
\begin{array}{cc}
\left(W^{(0)}\right)^{+k_0} & 0 \\
      0                     & \left(W^{(0)}\right)^{-k_0}
\end{array}
\right), \quad k_0 \in \Z,
\end{eqnarray}
with the unitary operator $W^{(0)}$ defined by Eq.~(\ref{W0}) and
the definition
\begin{eqnarray}
\left(W^{(0)}\right)^{-|k_0|} \equiv
\id \; \delta_{0,k_0}  +
\left(W^{(0)}{}^\dagger \right)^{|k_0|}\; \left( 1-\delta_{0,k_0} \right),
\quad
\mathrm{for} \quad k_0 \in \Z.
\end{eqnarray}
The corresponding basis spinors $v_j^{(0)\,\prime}$ from Eq.~(\ref{vjn}),
with $\tilde U^{(0)}$ replaced by $\tilde U^{(0)}{}'$,
are again gauge-covariant.

A straightforward calculation gives then for
the change of the measure under a CPT transformation
\begin{eqnarray}
\!\!\det \left( \sum_l \mathcal{Q}^{(0)}_{kl}[U]\,
            \bar{\mathcal{Q}}^{(0)}_{lm}\right)
=
\det \left(- a^3 \sum_{\vec x} \psi^\dagger_k(\vec x)
           \left(W^{(0)}[U]\right)^{2\,k_0+1} \psi_m(\vec x)\right),
\end{eqnarray}
which may be compared with Eq.~(\ref{Q0Q0bar}).
The corresponding change of the effective action under a CPT transformation is
given by
\begin{eqnarray}\label{DeltaGammageneral}
\Delta \Gamma[U] = - (2\, k_0+1)\;  \ln \det \left(-
  a^3 \sum_{\vec x} \psi^\dagger_k(\vec x)\, W^{(0)}[U]\, \psi_m(\vec x)\right),
\end{eqnarray}
which replaces Eq.~(\ref{DeltaGamma})  in the main text.

\end{appendix}

\end{document}